\documentclass{article}

\usepackage{PRIMEarxiv}

\usepackage[utf8]{inputenc} 
\usepackage[T1]{fontenc}    
\usepackage{hyperref}       
\usepackage{url}            
\usepackage{booktabs}       
\usepackage{amsfonts}       
\usepackage{nicefrac}       
\usepackage{microtype}      
\usepackage{lipsum}
\usepackage{fancyhdr}       
\usepackage{graphicx}       
\usepackage{xcolor}
\usepackage{soul}

\graphicspath{{media/}}     

\usepackage{amsmath}
\usepackage{hyperref}

\title{Discriminating sensor activation in activity recognition within multi-occupancy environments based on nearby interaction
}

\author{
  Aurora Polo-Rodríguez, Javier Medina-Quero \\
  Department of Computer Science \\
  University of Jaén \\
  Jaén, Spain\\
  \texttt{\{apolo,jmquero\}@ujaen.es} \\
}

\begin{document}
\maketitle

\begin{abstract}
In this work, we present a computer model to discriminate sensor activation in multi-occupancy environments based on proximity interaction. Current proximity-based and indoor location methods allow estimation of the positions or areas where inhabitants carry out their daily human activities. The spatial-temporal relation between location and sensor activations is described in this work to generate a sensor interaction matrix for each inhabitant. This enables the use of classical HAR models to reduce the complexity of the multi-occupancy problem. A case study deployed with UWB and binary sensors is presented.  
\end{abstract}


\section{Introduction}

The number of elderly people has increased significantly in recent decades due to various factors, such as increased life expectancy and improved health services \cite{wpp2022}. As more people require care, there is also a need for more caregivers to be involved in this field, which means higher costs and workloads for caregivers. However, while a growing number of older people prefer to stay at home for as long as possible rather than go to a care home \cite{donnelly2016d}, most of the population cannot afford the expense of having a carer at home. 

In this sense, smart environments have emerged as a solution to help elderly people to live safely, comfortably and independently in their homes, while reducing the toll on healthcare systems. The combination and use of different types of sensors to recognise Activities of Daily Living (ALD) is becoming increasingly common \cite{businesswire_2021} and includes vision sensors, audio sensors, wearables, binary sensors, RFID, PIR... under sensor fusion approaches \cite{lupion2021dolars}. Human activity recognition (HAR) aims to identify the daily tasks which occur in a given environment. Depending on the complexity and the sensorization of the system, we are able to determine fine-grained or coarse-grained granularity in the activities that the user is performing, defined as interleaved activities \cite{quero2018real}.


Indoor location systems are the perfect option to enhance and support sensors that do not have the ability to identify who activated them \cite{hayward2022survey}, which is why recently the scientific community has begun to combine activity recognition with sensor-based positioning systems \cite{zhan2021mosen}. A number of different approaches have been proposed such as fingerprint-based methods or ultrasound, infrared, Wi-Fi (IEEE 802.11), Bluetooth, ZigBee, Ultrawide Band (UWB), etc. \cite{oguntala2018indoor} It is essential to adapt to the characteristics of the actual environments in which multi-occupancy may occur. In multi-occupancy environments, it makes sense to design systems that are capable of identifying which inhabitant is carrying out the action, and this can be made possible by knowing the proximity of inhabitant to sensors \cite{gao2019livetag, meng2018distances}, associating the activation of sensors with the person who is near that area. It enables relating the human-object
interactions based on user proximity and sensor activation.


Considering the wide number of options, recognition of daily activities in a multi-occupancy smart home is a challenging goal due to the limited capabilities of current devices and the complexity of human tasks. Added to this is the challenge of finding a set of sensors that meets the requirements of high autonomy and low invasiveness.

The remainder of this paper is organised as follows: studies related to HAR and indoor location tracking multi-occupancy in intelligent environments with multi-occupancy are presented in Section 2. Section 3 provides an in-depth description of the methods used. Section 4 describes the configuration used in our case study, which can be independently replicated by any user. Section 5 presents our conclusions and future work.

\section{Related works}
\label{sec:related-works}

Sensor-based human activity recognition (HAR) requires a network of connected devices with the ability to track the daily tasks which a person is performing. The data produced are time series with certain (and uncertain) values or state changes \cite{vrigkas2015review, kim2009human}. These data generate knowledge describing ADL, which can also raise serious privacy concerns \cite{yang2017survey}. Because of this, systems based on binary sensors and wearables have dominated in this sector, as they are considerably less intrusive \cite{abade2018non}. HAR in smart homes is a big challenge, as human activity comprises complex tasks and can differ from day to day, just as it can change from inhabitant to inhabitant. Each person has different habits and capabilities.


Recent research on multi-occupancy environments has focused on the use of environmental, wearable and vision sensors. Several PIR (passive infrared) sensors located in the home are used in \cite{howedi2019exploring} to detect the presence of visitors in an unsupervised approach. The authors use data from the PIR sensors and various entropy measurements to establish a threshold that, when exceeded, is associated with the presence of a visitor. This nominal value is based on the standard deviation of the occupancy data together with various measures of entropy (namely Approximate Entropy, Sample Entropy, and Fuzzy Entropy). Several authors use similar approaches with PIR sensors to obtain information on occupancy in indoor spaces \cite{howedi2020employing, khan2021method, krishnamurthy2021determining}. 

However, most of the works are focused on detecting when there are visitors in a home where only one person lives, or attempts to identify the number of people in the environment, as it is difficult to associate the information received from environmental sensors to a specific individual. Vision sensors offer more accuracy in this regard, although this approach raises privacy \cite{dang2020sensor} and identification concerns. Recently, several works have addressed this issue using low-resolution thermal cameras \cite{zhong2020multi,polo2022classifying, manssor2021human,zhu2021privacy}. 


Generally, approaches to identify which person is performing an activity or activating a sensor are location-based, which is why several of the works described above include tracking people within the environment. Indoor location tracking remains a complex matter today, as current systems and devices often come up with false positive activations \cite{elhady2018systematic}. One of the first approaches to multi-occupancy indoor location was tackled by the authors of \cite{deak2014detection}, where they use Java SunSPOT nodes forming a wireless sensor network where the variation of RSSI (received signal strength indicator) intensity between nodes was monitored. Subsequently, a pattern recognition neural network was used to identify two people in the environment and detect which nodes they had passed through in a fully controlled room with almost no obstacles. Recently, authors have turned to voice-based systems using smart speakers, such as \cite{nath2018iot}, where Amazon Echo is used as a voice interface and an ultrasonic sensor for location detection and patient tracking. However, it is more common to use more widespread devices, such as smartphones. Some works use the RSSI values obtained from these devices, but the main problem is that the signals are highly variable from one model to another, even when they use the same operating system \cite{rizk2021device}.  Sensor fusion is a very useful approach to avoid such problems, as presented in \cite{liang2015smartphone}. The authors use the accelerometer and compass of an iPhone for step counting, and a location system based on a Kalman filter and RSSI values.

Moreover, the most widespread method of indoor location is the use of BLE due to its low cost and ease of deployment. Both \cite{tacsbacs2019real} and \cite{bai2020low} use BLE beacons and scanners together with fingerprinting. The difference is that the second proposal combines fingerprinting results with triangulation results obtained from RSSI. However, it obtains area accuracy (kitchen, bedroom, etc.) most of the time, without being able to distinguish in which area of the room the subject is located. In addition, the variability of the BLE signal and the presence of obstacles require installing a high density of beacons. Finally, the authors of \cite{tabbakha2021wearable} put forward a proposal closer to ours, which combines indoor location tracking and activity recognition with machine learning. In this case, they also rely on BLE through a wearable device on the waist, beacons and scanners to measure RSSI levels. They use random forest and support vector machine (SVM) as activity and location classifiers. However, the problems with BLE systems are again made apparent in this work, in addition to the fact that the recognised activities are limited: sitting, lying down, walking, falling and standing. Depending on the area where the subject is, it is associated with a sub-activity: for example, if they are standing in the kitchen, their activity is identified as cooking.

Based on all these proposals, we present our work in the following sections.

\section{Methods}

\begin{figure}[!ht]
\centering
    \includegraphics[width=0.8\linewidth]{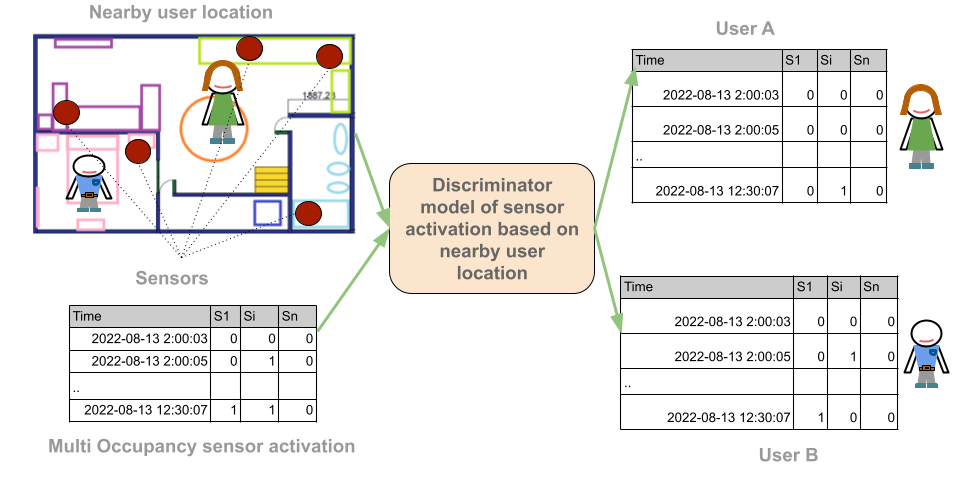}
    \caption{Description of inputs and outputs of the sensor activation and user proximity-based discriminator model}
    \label{fig:arq}
\end{figure}

In this section, we provide a formal description of our discriminator model based on sensor activation and proximity location. In Figure \ref{fig:arq}, we detail the data and components involved in the discriminator model, which is further described below. 

\subsection{Sensor activation and proximity location}

The inputs of the proposed models are based on sensing human-object interaction in smart environments.

First, a sensor $s$ is represented according to its activation in the form of pairs $\overline{s_i} =\{s_i,t_i\}$, where $s_i$ represents the given measurement of activation and $t_i$ is its timestamp. For the sake of simplicity, in this work we assume $s_i \in [0,1]$. In our case study, binary sensors are used where $s_i \in {0,1}$. The sensor $s$ generates a data stream which is defined by $\overline{S_s}=\{\overline{s_0},\ldots, \overline{s_i}\}$ where for a given timestamp $t_i$ we obtain the activation $S_s(t_i)=s_i$.

Second, location and proximity devices estimate the given location area $u_i$ of a user $u$ in a timestamp $t_i$. We note the location area $u_i$ is generally defined by geometrical space, such as, a point a, a circle or a bounding-box (rectangle). In our case study, we provided a point defined by $u_i=(x_i,y_i)$. Analogously, the user $u$ location is represented by a pair $\overline{u_i} =\{u_i,t_i\}$ generating a location stream $\overline{U_u}=\{\overline{u_0},\ldots, \overline{u_i}\}$.

\subsection{Temporal data segmentation}
\label{Temporal data segmentation}

Segmentation is a key step in HAR to determine the time-step decision-making, temporal alignment and data aggregation, and reduce computation complexity in sensor data processing. In this work, the time-step $\Delta$ and a given aggregation function $\bigcup$ configures the aggregation of the sensor and location stream:

\begin{equation}
\begin{aligned}
\overline{U_u}^*=\{\overline{u_0}^*,\ldots  \overline{u_i}^*\} \\
\overline{u_i}^*= \{u_i^*,t_i^*\} \\
t_i^*= \Delta \cdot i \\
u_i^*= \bigcup_{t_j}^{t_j \in [t_j^*, t_{j+1}^*)} u_j  \\
\end{aligned}
\end{equation}

For the sake of simplicity, we use $\overline{U_u}$ and $\overline{S_s}$ to refer to the aggregated and segmented data of the streams $\overline{S_s}^*$ and $\overline{U_u}^*$, respectively.

In our case study, we have defined the following aggregation functions:
\begin{itemize}
    \item Binary sensor activation $\bigcup=max(u_0, \ldots, u_i)$, whose semantic of activation is related to any binary activation of the sensor $s_j$ of the time-step which is defined within the temporal window $[t_j^*, t_{j+1}^*)$.
    \item Bounding-box proximity location, whose semantic is related to compose a bounding box defined by top left and bottom right points $u_i^*=\{(x_i^-,y_i^-),(x_i^+,y_i^+)\}$ from the location of the user within  the temporal window $\bigcup=\{(min(x_0, \ldots, x_i),min(y_0, \ldots, y_i)), (max(x_0, \ldots, x_i), max(y_0, \ldots, y_i))\}$.
\end{itemize}

\begin{figure}[!ht]
\centering
    \includegraphics[width=0.8\linewidth]{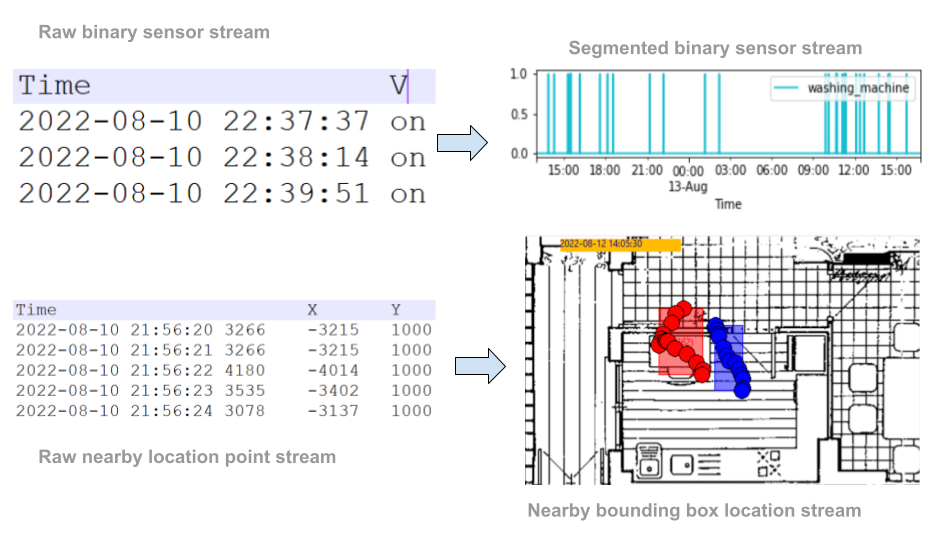}
    \caption{Segmentation and aggregation of binary sensor activation and user locations in time-steps}
    \label{fig:agg}
\end{figure}

In Figure \ref{fig:agg}, we describe an example of segmentation and aggregation of binary sensor activation and user locations in time-steps from raw sensor streams with $\Delta=15s$.

\subsection{Relating spatial-temporal interaction between inhabitants and sensors}
\label{Relating spatial-temporal interaction between inhabitants and sensors}

In this section, we detail the methods used to relate user location with sensor activation. The key aim is to compute the degree of interaction based on spatial location at the same point in time.

\begin{figure}[!ht]
\centering
    \includegraphics[width=0.8\linewidth]{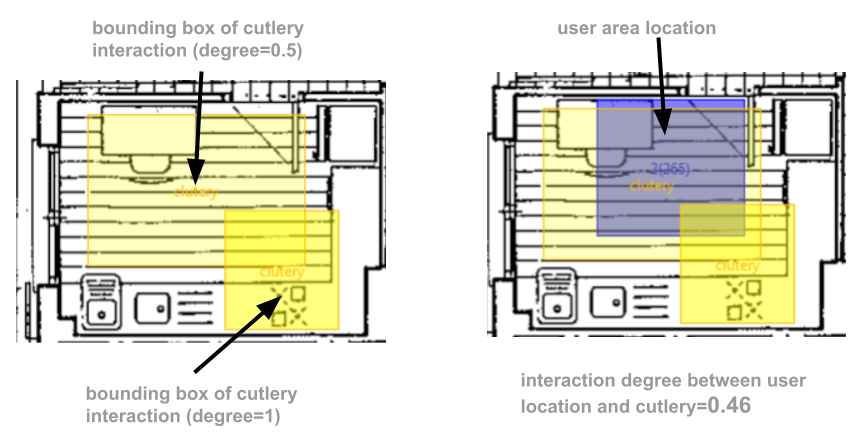}
    \caption{Left) example of interaction location areas for cutlery, and right) example of interaction areas between user location and cutlery}
    \label{fig:sensor_loc}
\end{figure}

First, we determine several interaction location areas $L(s)=\{l_0^u,\ldots, l_i^s\}$ for each sensor $s$ which is described by a interaction degree $\mu_{l_i^s}$. The aim of this degree is to describe several potential interaction areas with different interaction degrees to handle uncertainty and imprecision in user location. In Figure \ref{fig:sensor_loc}, we detail an example of interaction location areas for cutlery.

As previously indicated, areas are generally defined by a geometrical space, which must provide an intersection property, whose value between $[0,1]$ is defined as the interaction degree $\bigcap:L(s) \times \overline{U_u} \rightarrow [0,1]$ between the user location areas $\overline{U_u}$ and sensor location areas $L(s)$. For the sake of simplicity, we refer to the intersection as $l_j^s \bigcap u_i$. 

As several interaction location areas $l_j^s$ are defined for each sensor $s$, we compute the final interaction degree between the sensor $s$ and user location $u_i$ in the time stamp $t_i$ using the aggregation of the interaction location areas $\bigcup_{l_j^s}^{L(s)}$ weighted by their interaction degree $\mu_{l_j^s}$ as:

\begin{equation}
\begin{aligned}
L(s) \bigcap u_i = \bigcup_{l_j^s}^{l_j^s \in L(s)} (\mu_{l_j^s} \otimes l_j^s \bigcap u_i)
\end{aligned}
\end{equation}

In the case study described in this work, we present the following:
\begin{itemize}
    \item Sensor location areas $l_i^s$ are defined by a bounding box, as the same geometrical space as the aggregated locations of the users.
    \item The intersection between sensor location areas and user areas $l_j^s \bigcap u_i$ is defined from a modified version from the Jaccard index $J(A,B)=\frac{A \cap B}{A \cup B}$, which is also well known as intersection over union, and it is here modified as $J(l_j^s ,u_i)=\frac{l_j^s \bigcap u_i}{u_i}$ in order to only take into account the user area developed by the inhabitant track. In Figure \ref{fig:sensor_loc}, we show an example of interaction areas and computed degree between user location and cutlery.
    \item To weight intersection between sensor location areas and user areas with the interaction degree of the sensor areas $\mu_{l_j^s}$, we determine the product t-norm $a \otimes b= a \cdot b$-
    \item The aggregation of interaction location areas $\overset{l_j^s \in L(s)}{\underset{l_j^s}{\bigcup}}$ is computed by the Lukasiewicz t-norm $luk(a,b)=min(1,1-a+b)$, resulting in:
    
\begin{equation}
\begin{aligned}
L(s) \bigcap u_i =  \overset{l_j^s \in L(s)}{\underset{l_j^s}{luk}} ( \mu_{l_j^s} \cdot \frac{l_j^s \bigcap u_i}{u_i})
\end{aligned}
\end{equation}

\end{itemize}

\subsection{Semantics for discriminating sensor activation based on proximity interactions between sensors and inhabitants}

In this section, we describe the methods used to relate: i) the interaction degree $L(s) \bigcap u_i$, which describes the proximity relation  between the sensor $s$ and user location $u_i$, ii) the point in time $t_i$ where a given sensor $s$ is activated $s_i>0$.   

The interaction degree $L(s_i>0) \bigcap u_i$ for each user $u$ determines the spatial-temporal relation with the event of activation $s_i>0$ for the sensor $s$. If the degree is $L(s_i>0) \bigcap u_i>0$, we note there is a spatial-temporal relation with the user, and no relation otherwise $L(s_i>0) \bigcap u_i=0$. 

The aim of this work is to relate the data stream for the sensor $s$ which is defined by $\overline{S_s}$ with a given user $u$, whose location is defined by $\overline{U_u}$, in order to discriminate the user-sensor interaction degree of a sensor stream for each user $\overline{S_s^u}=\{\overline{s_0^u},\ldots, \overline{s_i^u}\}$.

At the end of this section, we describe several semantics to determine the activation for each user based on different proximity-based discrimination policies:

\begin{itemize}
    \item Multiple user interaction, where several users may have interacted with the sensor. In this case, the interaction degree $L(s_i>0) \bigcap u_i$  provides an straightforward semantic for describing the user-sensor interaction degree: $\overline{s_i^u}=L(s_i>0) \bigcap u_i$, where several users are able to interact with same sensor and in the same time-step $\overline{s_i^A}>0, \overline{s_i^B}>0, A \neq B $.

    \item Exclusively single-user interaction, where only one user may have interacted with the sensor. In this case, the interaction is uniquely related to a higher user degree defined by the operator $sup$, which is represented as maximum in this work :
    
\begin{equation}
\begin{aligned}
\overline{s_i^u} =  \left\{ \begin{array}{lcc}
             L(s_i>0) \bigcap u_j & if & sup(L(s) \bigcap u_j),  \forall u_j \in U \\
             \\ 0 & & otherwise
             \end{array}
   \right.
\end{aligned}
\end{equation}

    \item Others. We note the use of other methods to compute the interaction degree between inhabitants and objects based on proximity, such as low filter or fuzzy quantification to increase the representation of the semantic interaction.  
    
\end{itemize}

\section{Case study}

In this section, we describe the case study developed in this work to evaluate and present the use of the methodology in a real-life context. The case study was developed in a kitchen of a flat where 4 inhabitants (2 adults and 2 children) carry out their daily activities in an ordinary living space. The location of the two adults and the opening/closing of 4 home appliances/furniture (cutlery, dishwasher, fridge, microwave) was monitored using UWB (ultra-wideband) and binary sensors, respectively.

The data, method implementation and results of this work have been made openly available at \url{https://github.com/AuroraPR/HAR-nearby-interaction}.

\subsection{Installation of sensors}

\begin{figure}[!ht]
\centering
    \includegraphics[width=0.8\linewidth]{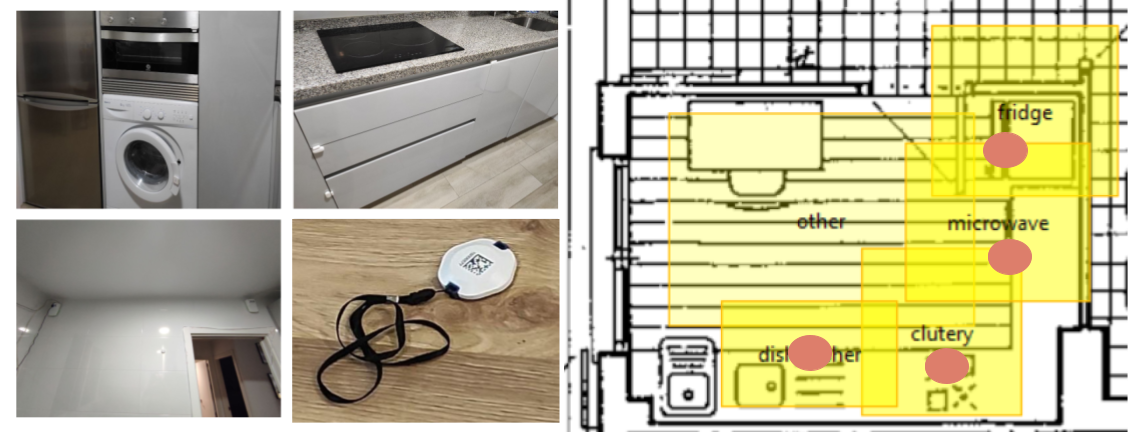}
    \caption{Left) deployment in a real kitchen environment: binary, acceleration and UWB anchors and tag. Right)  interaction location areas related to sensors installed on home appliances/furniture}
    \label{fig:deployment_sensors}
\end{figure}

First, the location of the inhabitants was tracked by the proximity sensor technology \url{https://www.pozyx.io} based on UWB. Four anchors were located in the corners of the kitchen ceiling to provide the location (x,y) of the inhabitants, who wore a UWB wearable tag (see Figure \ref{fig:deployment_sensors})). The platform provides trilaterion estimation in real time using the MQTT protocol.

Second, opening/closing events by the inhabitants in the home appliances/furniture of the kitchen were detected using Xiaomi Aqara binary and acceleration sensors \url{https://www.aqara.com/}. The binary sensors were integrated in the Home Assistant platform \url{https://www.home-assistant.io/}, which provides real-time connection by MQTT. Each sensor was placed on an appliance/furniture item (see location as red points in Figure \ref{fig:deployment_sensors}).

Based on this proximity sensor deployment, we collected data for one day (two uses for cooking lunch and daily events in the afternoon) in a real-life kitchen use scenario. In Figure \ref{fig:deployment_sensors}, we show pictures of the deployment and UWB trilateral positioning. 

\subsection{Parameter setup}

In this section, we show the configuration of the interaction location areas related to sensors installed on the home appliances/furniture deployed in the case study.  Each home appliance/furniture item determines a closer interaction area, which is defined as degree=1, as well as an external common area for working in the kitchen (other), which is defined as degree=0 (see Figure \ref{fig:deployment_sensors}).

The interaction policy related to the sensors is: multiple user interaction for fridge and dishwasher, and exclusively single-user interaction for cutlery and microwave.

The time-step was set to $\Delta=15s$, which determines the aggregation of locations in the bounding-box areas and the coarse-granularity of activation of binary sensors.

\subsection{Results}

In this section, we describe the results obtained in the case study. First, we describe the binary sensor activation deployed on the home appliances/furniture corresponding to daily human activities. They were activated by the 4 inhabitants (only 2 adults were monitored by location tracking of UWB wearable tag), so the activation is mainly triggered by sensorized inhabitants (adults) and  by others (children).

\begin{figure}[!ht]
\centering
    \includegraphics[width=0.8\linewidth]{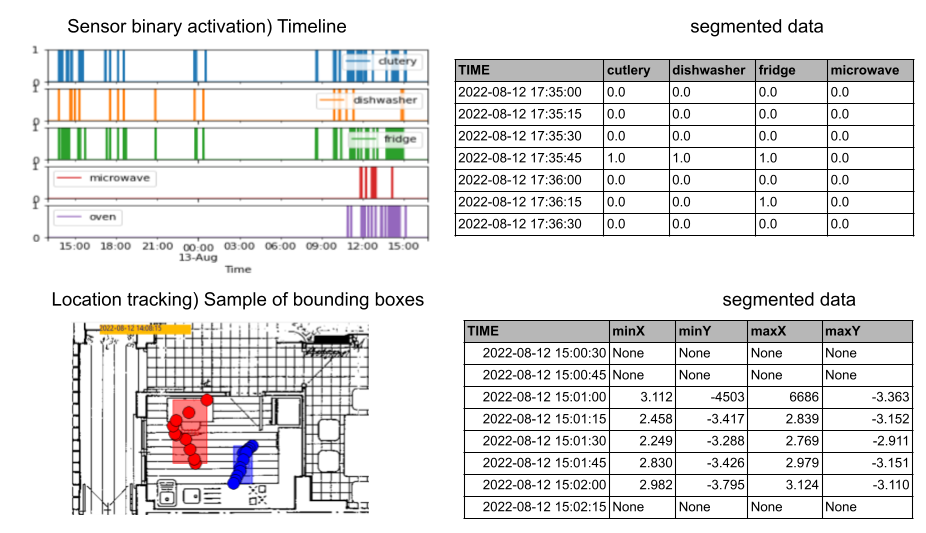}
    \caption{Top) Example of segmented sensor binary activation; Bottom) Example of segmented tracking location}
    \label{fig:segmented-data}
\end{figure}

The number of raw events/activations collected by the sensors is presented in Table \ref{table:activation}, and the timeline in Figure \ref{fig:activation}. In addition, the number of inhabitant locations sensed by UWB is included in Table \ref{table:activation}. The number of instances by the segmentation/aggregation method described in Section \ref{Temporal data segmentation} is likewise included in Table \ref{table:activation} and shown in Figure \ref{fig:segmented-data}. We note the significant reduction of complexity in the size of the data from high-rate sensors thanks to the UWB location technology.

\begin{table} [!ht]

\centering
\caption{Number of events collected by binary sensor activation and tracking location of inhabitants in the case study (raw and segmented).}
\label{table:activation}
{\small
  \begin{tabular}{| p{3.2cm} | p{1.0cm}| p{1.0cm}| p{1.0cm}| p{1.0cm}| p{1.0cm}| p{1.0cm}|}
    \hline
     Instances & \multicolumn{4}{|c|}{Sensor activation} &
      \multicolumn{2}{|c|}{Location tracking} \\ \hline
    & \scriptsize{cutlery} & \scriptsize{dishwasher} & \scriptsize{fridge} & \scriptsize{microwave} & \scriptsize{user A} & \scriptsize{user B}  \\ \hline
        Raw & \scriptsize{136}  & \scriptsize{26} & 
        \scriptsize{172} & \scriptsize{12}  & \scriptsize{60081} & \scriptsize{46989}  \\  \hline        
        Segmented $\Delta=15s$ & \scriptsize{98}  & \scriptsize{24} & 
        \scriptsize{136} & \scriptsize{11}  & \scriptsize{1818} & \scriptsize{1664}  \\  \hline          
  \end{tabular}}
\end{table}

\begin{figure}[!ht]
\centering
    \includegraphics[width=0.8\linewidth]{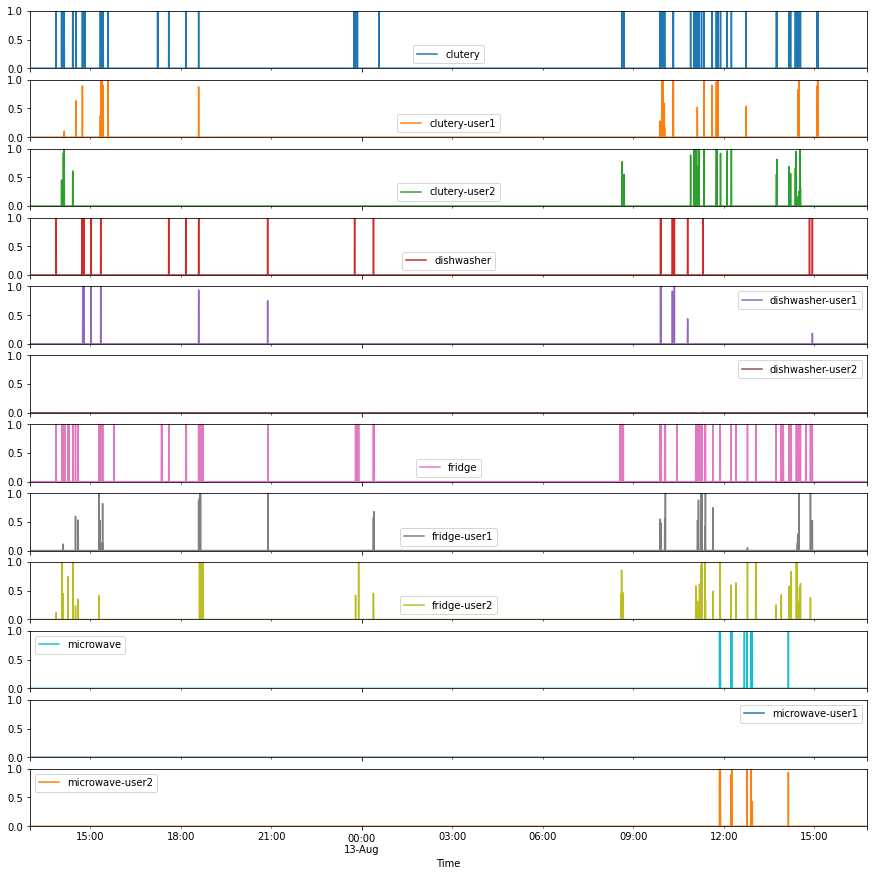}
    \caption{For each sensor activation, we have raw activation, interaction degree of user A and interaction degree of user B, which is computed from the spatial-temporal discrimination method.}
    \label{fig:res_evaluation_metrics}
\end{figure}

At the end, we computed the discrimination of sensor activation based on the proximity interactions of inhabitants as set out in Section \ref{Relating spatial-temporal interaction between inhabitants and sensors}. The spatial-temporal discrimination segments the data into user A and B according to the sensor activation described in Table \ref{tab:res_evaluation_metrics}. In this case study, we note the method also indicates events which are not provided by tracked users, such as children. In addition, we observe differences in the behaviour patterns of the inhabitants: the dishwasher is mainly used by inhabitant A who takes on more of a cleaning role; meanwhile, the fridge/microwave is used mainly by inhabitant B who takes on more of a cooking role. In Figure \ref{fig:res_evaluation_metrics}, we show the timeline of raw and discriminated activation data collected for the inhabitants in the case study.

\begin{table}[]
\centering
\begin{tabular}{|c|c|c|c|c|}
\hline
\scriptsize{Inhabitant}  & \scriptsize{cutlery} & \scriptsize{dishwasher} & \scriptsize{fridge} & \scriptsize{microwave}  \\ \hline
user A &	27	& 9 &	38 &	0 \\ \hline
user B &	30 &	1 &	68 &	3 \\ \hline
others &	41 & 14 & 30 &		8 \\ \hline
Total &	98 &	24 &	136 &	11 \\ \hline

\end{tabular}
\caption{Spatial-temporal discrimination of sensor activation for user A and B.}
\label{tab:res_evaluation_metrics}
\end{table}

\section{Conclusion}

In this work, we describe a computer model that discriminates the activation of binary sensors in multi-occupancy environments in order to distinguish which user performed each action based on proximity. The challenge has been presented together with related works and a formal description of the sensors and methods proposed to solve the problem.

A case study based on Pozyx UWB and Aquara sensors has been deployed and evaluated with interesting results regarding the roles of the two inhabitants involved.

In ongoing works, we will collect a wide multi-occupancy dataset from several homes to provide more comprehensive results. In addition, we will evaluate and discriminate interactions based on the time-intervals, not only sensor activations in points of time.

\section*{Acknowledgments}
This contribution has been supported by the Spanish Institute of Health ISCIII by means of project DTS21-00047. 

\section*{License}
This work is under licence of the register
\textit{MERCEDES 2.0: Activity recognition system in multi-occupancy environments/MERCEDES 2.0: Sistema de reconocimiento de actividades en entornos de multiocupación}, with proof of
ID in Safe Creative: 2211022548670.

\bibliographystyle{unsrt}  
\bibliography{references}

\begin{thebibliography}{10}

\bibitem{wpp2022}
World population prospects - population division.

\bibitem{donnelly2016d}
Sarah Donnelly, Marita O'Brien, Emer Begley, and John Brennan.
\newblock 'i'd prefer to stay at home but i don't have a choice': Meeting older
  people's preference for care: Policy, but what about practice?
\newblock Technical report, University College Dublin. School of Social Policy,
  Social Work and Social~…, 2016.

\bibitem{businesswire_2021}
Global smart home market research report (2021 to 2026) - by product,
  technologies, service, deployment and region - researchandmarkets.com, Dec
  2021.

\bibitem{lupion2021dolars}
Marcos Lupi{\'o}n, Javier Medina-Quero, Juan~F Sanjuan, and Pilar~M Ortigosa.
\newblock Dolars, a distributed on-line activity recognition system by means of
  heterogeneous sensors in real-life deployments—a case study in the smart
  lab of the university of almer{\'\i}a.
\newblock {\em Sensors}, 21(2):405, 2021.

\bibitem{quero2018real}
Javier~Medina Quero, Claire Orr, Shuai Zang, Chris Nugent, Alberto Salguero,
  and Macarena Espinilla.
\newblock Real-time recognition of interleaved activities based on ensemble
  classifier of long short-term memory with fuzzy temporal windows.
\newblock {\em Multidisciplinary digital publishing institute proceedings},
  2(19):1225, 2018.

\bibitem{hayward2022survey}
SJ~Hayward, Kate van Lopik, Christopher Hinde, and AA~West.
\newblock A survey of indoor location technologies, techniques and applications
  in industry.
\newblock {\em Internet of Things}, page 100608, 2022.

\bibitem{zhan2021mosen}
Yuting Zhan and Hamed Haddadi.
\newblock Mosen: Activity modelling in multiple-occupancy smart homes.
\newblock {\em arXiv preprint arXiv:2101.00235}, 2021.

\bibitem{oguntala2018indoor}
George Oguntala, Raed Abd-Alhameed, Stephen Jones, James Noras, Mohammad
  Patwary, and Jonathan Rodriguez.
\newblock Indoor location identification technologies for real-time iot-based
  applications: An inclusive survey.
\newblock {\em Computer Science Review}, 30:55--79, 2018.

\bibitem{gao2019livetag}
Chuhan Gao, Yilong Li, and Xinyu Zhang.
\newblock Livetag: Sensing human-object interaction through passive chipless
  wi-fi tags.
\newblock {\em GetMobile: Mobile Computing and Communications}, 22(3):32--35,
  2019.

\bibitem{meng2018distances}
Meng Meng, Hassen Drira, and Jacques Boonaert.
\newblock Distances evolution analysis for online and off-line human object
  interaction recognition.
\newblock {\em Image and Vision Computing}, 70:32--45, 2018.

\bibitem{vrigkas2015review}
Michalis Vrigkas, Christophoros Nikou, and Ioannis~A Kakadiaris.
\newblock A review of human activity recognition methods.
\newblock {\em Frontiers in Robotics and AI}, 2:28, 2015.

\bibitem{kim2009human}
Eunju Kim, Sumi Helal, and Diane Cook.
\newblock Human activity recognition and pattern discovery.
\newblock {\em IEEE pervasive computing}, 9(1):48--53, 2009.

\bibitem{yang2017survey}
Yuchen Yang, Longfei Wu, Guisheng Yin, Lijie Li, and Hongbin Zhao.
\newblock A survey on security and privacy issues in internet-of-things.
\newblock {\em IEEE Internet of Things Journal}, 4(5):1250--1258, 2017.

\bibitem{abade2018non}
Bruno Abade, David Perez~Abreu, and Marilia Curado.
\newblock A non-intrusive approach for indoor occupancy detection in smart
  environments.
\newblock {\em Sensors}, 18(11):3953, 2018.

\bibitem{howedi2019exploring}
Aadel Howedi, Ahmad Lotfi, and Amir Pourabdollah.
\newblock Exploring entropy measurements to identify multi-occupancy in
  activities of daily living.
\newblock {\em Entropy}, 21(4):416, 2019.

\bibitem{howedi2020employing}
Aadel Howedi, Ahmad Lotfi, and Amir Pourabdollah.
\newblock Employing entropy measures to identify visitors in multi-occupancy
  environments.
\newblock {\em Journal of Ambient Intelligence and Humanized Computing}, pages
  1--14, 2020.

\bibitem{khan2021method}
Donya~Sheikh Khan, Jakub Kolarik, Christian~Anker Hviid, and Peter Weitzmann.
\newblock Method for long-term mapping of occupancy patterns in open-plan and
  single office spaces by using passive-infrared (pir) sensors mounted below
  desks.
\newblock {\em Energy and Buildings}, 230:110534, 2021.

\bibitem{krishnamurthy2021determining}
Raj Krishnamurthy.
\newblock Determining occupancy of a multi-occupancy space, March~30 2021.
\newblock US Patent 10,963,683.

\bibitem{dang2020sensor}
L~Minh Dang, Kyungbok Min, Hanxiang Wang, Md~Jalil Piran, Cheol~Hee Lee, and
  Hyeonjoon Moon.
\newblock Sensor-based and vision-based human activity recognition: A
  comprehensive survey.
\newblock {\em Pattern Recognition}, 108:107561, 2020.

\bibitem{zhong2020multi}
Cankun Zhong, Wing~WY Ng, Shuai Zhang, Chris~D Nugent, Colin Shewell, and
  Javier Medina-Quero.
\newblock Multi-occupancy fall detection using non-invasive thermal vision
  sensor.
\newblock {\em IEEE Sensors Journal}, 21(4):5377--5388, 2020.

\bibitem{polo2022classifying}
Aurora Polo-Rodriguez, Alicia Montoro-Lendinez, Macarena Espinilla, and Javier
  Medina-Quero.
\newblock Classifying sport-related human activity from thermal vision sensors
  using cnn and lstm.
\newblock In {\em International Conference on Image Analysis and Processing},
  pages 38--48. Springer, 2022.

\bibitem{manssor2021human}
Samah~AF Manssor, Zhengyun Ren, Rong Huang, and Shaoyuan Sun.
\newblock Human activity recognition in thermal infrared imaging based on deep
  recurrent neural networks.
\newblock In {\em 2021 14th International Congress on Image and Signal
  Processing, BioMedical Engineering and Informatics (CISP-BMEI)}, pages 1--7.
  IEEE, 2021.

\bibitem{zhu2021privacy}
Shuai Zhu.
\newblock Privacy-preserving building occupancy estimation via low-resolution
  infrared thermal cameras, 2021.

\bibitem{elhady2018systematic}
Nancy~E ElHady and Julien Provost.
\newblock A systematic survey on sensor failure detection and fault-tolerance
  in ambient assisted living.
\newblock {\em Sensors}, 18(7):1991, 2018.

\bibitem{deak2014detection}
Gabriel Deak, Kevin Curran, Joan Condell, and Daniel Deak.
\newblock Detection of multi-occupancy using device-free passive localisation.
\newblock {\em IET Wireless Sensor Systems}, 4(3):130--137, 2014.

\bibitem{nath2018iot}
Rajdeep~Kumar Nath, Rajnish Bajpai, and Himanshu Thapliyal.
\newblock Iot based indoor location detection system for smart home
  environment.
\newblock In {\em 2018 ieee international conference on consumer electronics
  (icce)}, pages 1--3. IEEE, 2018.

\bibitem{rizk2021device}
Hamada Rizk, Moustafa Abbas, and Moustafa Youssef.
\newblock Device-independent cellular-based indoor location tracking using deep
  learning.
\newblock {\em Pervasive and Mobile Computing}, 75:101420, 2021.

\bibitem{liang2015smartphone}
Po-Chou Liang and Paul Krause.
\newblock Smartphone-based real-time indoor location tracking with 1-m
  precision.
\newblock {\em IEEE journal of biomedical and health informatics},
  20(3):756--762, 2015.

\bibitem{tacsbacs2019real}
Ahmet~Semih Ta{\c{s}}ba{\c{s}}, Emre Erdal, and Suat {\"O}zdemir.
\newblock Real-time object and personnel tracking in indoor location.
\newblock In {\em 2019 4th International Conference on Computer Science and
  Engineering (UBMK)}, pages 585--590. IEEE, 2019.

\bibitem{bai2020low}
Lu~Bai, Fabio Ciravegna, Raymond Bond, and Maurice Mulvenna.
\newblock A low cost indoor positioning system using bluetooth low energy.
\newblock {\em Ieee Access}, 8:136858--136871, 2020.

\bibitem{tabbakha2021wearable}
Nour~Eddin Tabbakha, Chee~Pun Ooi, Wooi~Haw Tan, and Yi-Fei Tan.
\newblock A wearable device for machine learning based elderly's activity
  tracking and indoor location system.
\newblock {\em Bulletin of Electrical Engineering and Informatics},
  10(2):927--939, 2021.

\end{thebibliography}

\end{document}